\documentstyle[prl,aps,twocolumn,epsfig]{revtex}

\newcommand{\rr}{{\bf r}}
\newcommand{\kk}{{\bf k}}
\newcommand{\qq}{{\bf q}}
\newcommand{\pp}{{\bf p}}
\newcommand{\uu}{{\bf u}}
\newcommand{\vv}{{\bf v}}
\newcommand{\QQ}{{\bf Q}}

\newcommand{\ee}{{\mbox{e}}}
\newcommand{\ii}{{\mbox{\footnotesize i}}}
\newcommand{\II}{{\mbox{i}}}
\newcommand{\beq}{\begin{equation}}
\newcommand{\eeq}{\end{equation}}

\begin{document}

\author{Radka Bach$^{1}$, Marek Trippenbach$^{2}$ and Kazimierz Rz{\c{a}}{\.z}ewski$^{1}$ \\
   {\it $^1$ Center for Theoretical Physics, Polish Academy of Sciences, al. Lotnik{\'o}w 32/46,
02-668 Warsaw, Poland \\
$^2$ Institute of Experimental Physics, Optics Division, Warsaw University, 69 Ho{\.z}a Street, 00-681
Warsaw, Poland}}
\title{Spontaneous emission of atoms via collisions of Bose-Einstein condensates}

\maketitle

\begin{abstract}

The widely used Gross-Pitaevskii equation treats only coherent aspects of the evolution of a Bose-Einstein condensate.
However, inevitably some atoms scatter out of the condensate. We have developed a method, based on the field theory
formulation, describing the dynamics of incoherent processes which are due to elastic collisions. We can therefore treat
processes of spontaneous emission of atoms into the empty modes, as opposed to stimulated processes, which require
non-zero initial occupation.

In this article we study two counter-propagating plane waves of atoms, calculating the full dynamics of mode occupation, as
well as the statistics of scattered atoms. The more realistic case of Gaussian wavepackets is also analyzed.

\end{abstract}\hspace{0.5cm}

In recent years, there is a growing number of experiments in which the atomic Bose-Einstein condensate undergoes a
nontrivial dynamics. Some examples are bouncing condensates \cite{1}, dark solitons in a condensate \cite{2,3}, vortices
and their dynamics \cite{4}, Bragg splitting of the condensate \cite{5} and 
many others.

A remarkably universal tool describing all these experiments is the celebrated Gross-Pitaevskii equation (GPE)
\cite{Gross,Pitaev}. It describes a coherent evolution of the mean atomic field. In the most common interpretation, its
time dependent version assumes that each atom of the system undergoes identical evolution. This is a good assumption since
in these experiments, initially the wave-packet of the system contains thousands of particles. To use a term borrowed from
quantum optics, the time dependent GPE describes {\it stimulated processes}.  At first glance, atomic four wave mixing
looks like an exception. It is not, since in this case the nonlinear term in the GPE contains a macroscopic source term of
the fourth wave. The analogous optical process is also described well by the c-number version of Maxwell equations with a
nonlinear polarization \cite{7}. Similarly, the second harmonic generation requires no field quantization, but the reverse
process of the parametric down conversion starts-off due to vacuum fluctuations and requires quantized photons \cite{7}.
Also, the mechanism of spontaneous emission is necessary for single atom radiative decay and its multi-atom generalization
known as superfluorescence \cite{8}.

In the process of a collision between two condensates, inevitably, some atoms from both droplets of quantum matter would
scatter away from the two condensates. This is a loss mechanism which is not accounted for by the conventional GPE. The
models were developed in which the effects of this loss on the condensates dynamics were incorporated in the GPE within
the slowly varying envelope approximation technique \cite{9}. It is a purpose of this article to propose a simple method of
calculating not only these losses but also quantum statistical properties of the atoms scattered away. Our starting point
will be a second quantized version of the many-atom Hamiltonian. Taking our inspiration from quantum optics, we shall
approximate this Hamiltonian in such a way, that the only quntized modes of the atom field will be those which initially
are in the vacuum state. Solving the Heisenberg and Schr{\"o}dinger equations for the empty modes we shall be able to
answer all relevant questions concerning the lost atoms.

The Hamiltonian describing a multi-atom system moving in a large box of length L and undergoing cold collisions via a
contact potential reads
\begin{eqnarray}
H &=& \int_{V=L^3} d^3 r \Psi^\dagger(\rr,t) \frac{\hat{p}^2}{2m} \Psi(\rr,t) + \nonumber \\
  &+& \frac{g}{2} \int_{V=L^3} d^3 r \Psi^\dagger(\rr,t) \Psi^\dagger(\rr,t) \Psi(\rr,t) \Psi(\rr,t)
   \label{full:ham}
\end{eqnarray}
where $\Psi$ denotes an atomic field operator satisfying the standard equal time Bose commutation relations and coupling
constant $g = 4 \pi \hbar^2 a / m$, where $a$ is the s-wave scattering length.

In this article we consider a collision processes of two large Bose-Einstein condensates. In particular we want to compute
the part of the atomic field consisting of atoms scattered away from both condensates. In this process two out of four
field operators in the last term in (\ref{full:ham}) may be replaced by the c-number colliding wave-packets. In the
scattering process two atoms from these wave-packets are destroyed and two atoms appear in the initially empty modes. As
the losses are assumed small, in the crudest approximation we do not take into account the modification of the colliding
condensates due to the scattering process. In this way we obtain a Hamiltonian, which for the problem considered here
resembles the radiation of given current in quantum electrodynamics \cite{10} or a parametric down conversion with
undepleted pump in quantum optics \cite{9}. Its hermitian form contains both creation and annihilation terms:
\begin{eqnarray}
H &=& \int d^3 r \Psi_e^\dagger(\rr,t) \frac{\hat{p}^2}{2m} \Psi_e(\rr,t) + \nonumber \\
  &+& \frac{g}{2} \int d^3 r \Psi_e^\dagger(\rr,t) \Psi_e^\dagger(\rr,t)
\psi_{\QQ}(\rr,t) \psi_{-\QQ}(\rr,t) + h.c.
\label{ham}
\end{eqnarray}
Here $\psi_{\pm \QQ}$ are the wave functions of colliding condensates (in the center of mass reference frame) and $\Psi_e$
is the field operator of the empty modes. Hamiltonian (\ref{ham}) is quadratic in the atomic field operators. Hence, the
field equations will be linear and tractable. We pay a price for this dramatic simplification: unlike (\ref{full:ham}) the
Hamiltonian (\ref{ham}) does not conserve the number of particles. Also the dynamics of the condansates cannot be found in
this approach, GPE must be solved beforehand.

The roots of this approximation, together with the validity conditions, might be traced back to the time-dependent
generalization of the Bogoliubov approach, in which the boson field operator is decomposed into the macroscopically
occupied part (i.e. the condensate) and the quantum corrections. As our interest lies in collision processes of
Bose-Einstein condensates, the wavefunction of the whole system is just the sum of colliding condensate's wavefunctions.
The Bogoliubov decomposition can be therefore written as:
\begin{equation}
   \hat{\Psi}(\rr,t)= \psi_\QQ(\rr,t) + \psi_{-\QQ}(\rr,t) + \hat{\Psi}_e(\rr,t)
   \label{Bog}
\end{equation}
Inserting (\ref{Bog}) into the Hamiltonian (\ref{full:ham}) one obtains a collection of different terms, of which only two
remain in Hamiltonian (\ref{ham}). We shall argue that such an approximation gives correct results if and only if the
kinetic energy associated with the center of mass motion is much larger than the interaction energy per particle, $\hbar^2
Q^2 / (2m) \gg g n$, where $n$ is the average density of the particles in the condensates. 
This condition is readily fulfilled in current experiments (see for example \cite{Philips,experiment,Ketterle}).

\begin{figure} \centering
\renewcommand{\arraystretch}{0.}
\begin{tabular}{c}
\epsfxsize=5.5cm
\epsffile{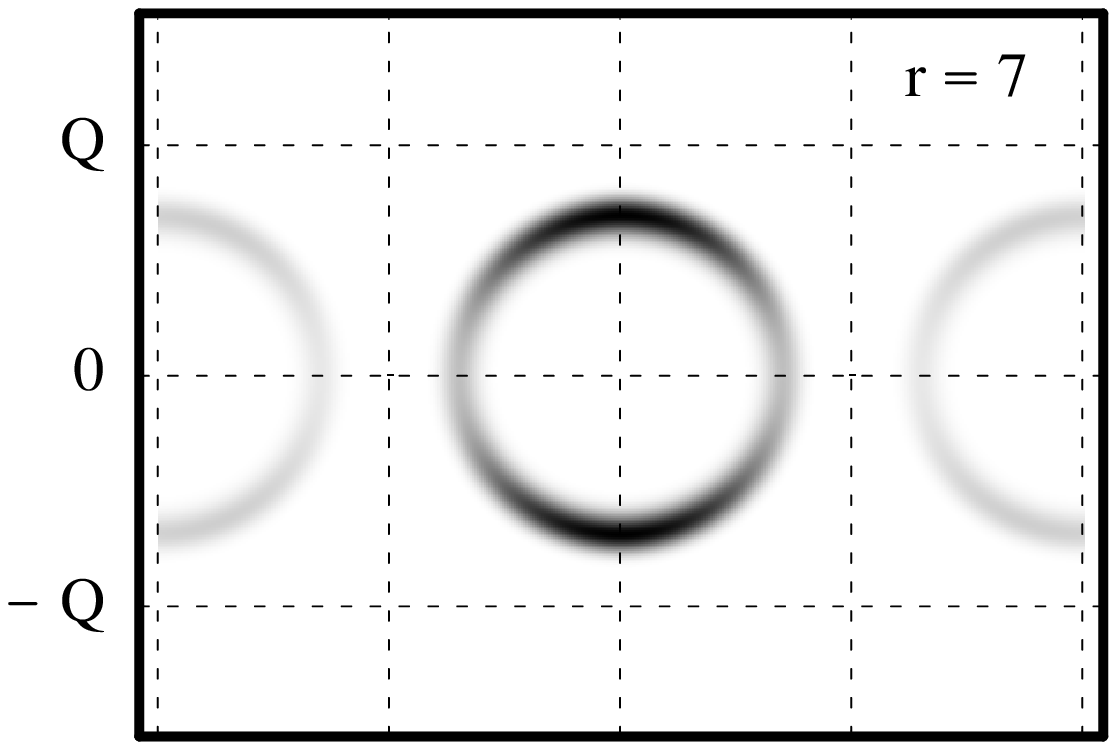} \\
\epsfxsize=5.5cm
\epsffile{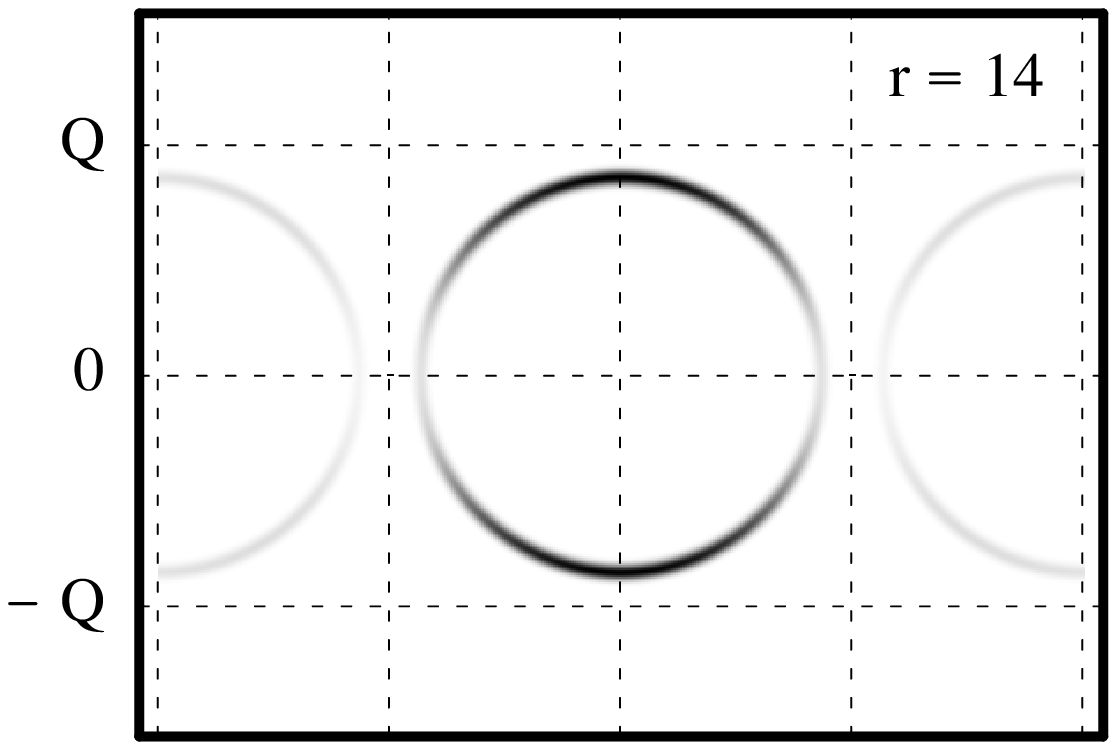} \\
\epsfxsize=5.5cm
\renewcommand{\arraystretch}{1.}
\epsffile{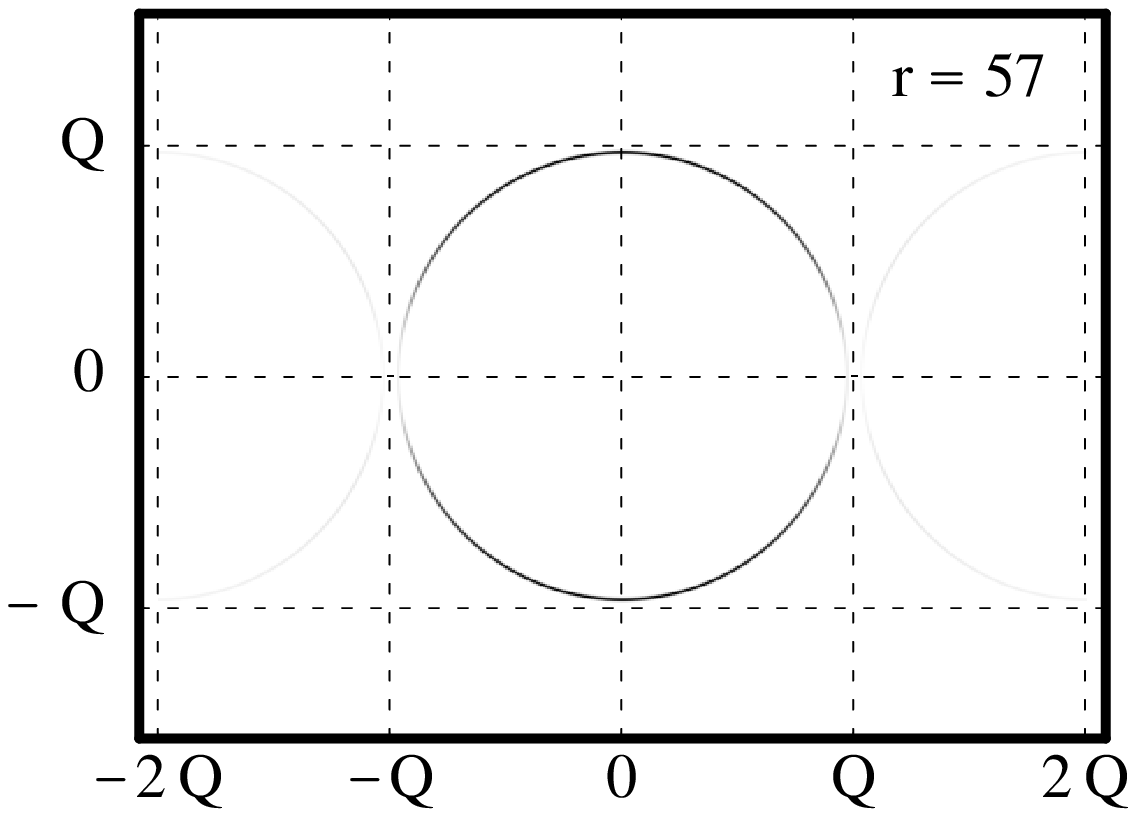} \\
\phantom{ala}\\
\end{tabular}
\caption{Bogoliubov approximation: population of modes to the power $1/3$ 
   (to make secondary rings visible) on a momentum plane. Two
initial plane waves are located at $(Q,0)$ and $(-Q,0)$. The parameter $r$ is the ratio of the kinetic energy
associated with the center of mass motion to the interaction energy per particle $r=\frac{\hbar^2 Q^2}{2 m g n}$.
Large $r$ are well described by our approximation.}
\label{rys:kolka}
\end{figure}

We can prove the above statement in the simplest case of two counter-propagating plane matter waves. 
The wavefunctions of the condensates evolving according to the free Hamiltonian are the following:
\beq \psi_{\pm \QQ}(\rr,t) = \sqrt{n_{\pm \QQ}} \: \ee^{\mp \ii \QQ \rr } \: \ee^{-\ii \hbar \frac{Q^2}{2m} t} \label{3} \eeq 
where $n_{\pm \QQ}$ is the density of
particles. Let us assume for simplicity that both waves are equally populated, $n_{\QQ}=n_{-\QQ}=n$. In a box
of length $L$, the boson field operator can be decomposed into normalized plane waves: \beq \Psi_e(\rr,t) =
L^{-3/2} \sum_{\qq} e^{-i \qq \rr} a_{\qq}(t) \label{r:3} \eeq where $a_{\qq}(t)$ is an annihilation
operator of an atom of the wavevector $\qq$. Bogoliubov approximation implies the following:
\begin{eqnarray}
\dot{a}_\pp &=& - \II \hbar \frac{p^2}{2m} a_\pp 
    - 2 \II \frac{gn}{\hbar} \left(  a_{\pp-2\QQ} + 2 a_{\pp} + a_{\pp+2\QQ}  \right) + \nonumber \\
  && - \II \frac{gn}{\hbar} \ee^{- \ii \hbar Q^2 t /m} 
\left(  a_{2\QQ-\pp}^\dagger + 2  a^\dagger_{-\pp} + a^\dagger_{-2\QQ-\pp} \right)
   \label{r:Bog}
\end{eqnarray}
This way a set of coupled linear equations for creation/annihilation operators is obtained. To solve it we have
introduced a cut-off in the number of modes taken into account and 
then solved the resulting finite system algebraically. The numerically found populations of different modes 
are plotted in Fig.\ \ref{rys:kolka}; the figures differ by the value of total kinetic energy of
colliding waves. 

Different terms in Eq.\ (\ref{r:Bog}) result in different aspects of mode's population in Fig.\ \ref{rys:kolka}.
The terms proportional to $- \II \hbar \frac{p^2}{2m} a_\pp$ and $a_{-\pp}^\dagger$
(the only ones that remain in our approximation presented in this paper) are responsible for 
the elastic scattering of atoms from two opposite waves. They form 
the main circle centered at $p=0$ in momentum plane that appears on the graphs. 
Increasing the value of kinetic energy (increasing $r$) results in relative sharpening of the circle $p\approx Q$.
One can see that for large values of $Q$ populating this circle is the main effect of spontaneous emission.

The other terms, $- 2 \II \frac{gn}{\hbar} \left(  a_{\pp-2\QQ} + 2 a_{\pp} + a_{\pp+2\QQ}  \right)$
are responsible for the frequency shifts, i.e. the fact, that the circles' radii are smaller than $Q$. 
Due to these terms also anisotrophy in the circle population appears (scattering perpendicular to the axis of the
colliding waves is the most probable).
However, these effects are getting negligible as $Q$ increases -- the circles grow, finally reaching $p= Q$ and
flatten.

The last terms, $- \II \frac{gn}{\hbar} \ee^{- \ii \hbar Q^2 t /m}
\left(  a_{2\QQ-\pp}^\dagger + a^\dagger_{-2\QQ-\pp} \right)$, are responsible for secondary rings, which are a result of
processes in which two atoms from the same matter wave scatter. After the collision one of them is placed in 
the main circle (because of the Bose enhancement) and the other must go to the secondary ring to fulfill the momentum
conservation. However, the kinetic energy in such a process is not conserved: the energy of atoms after the collision
is greater than before; the difference is taken from the interaction energy. Thus, as the ratio of kinetic energy to
the interaction energy increases (parameter $r$) these processes are less and less important and finally practically
disappear for large $Q$.

Hence we have shown that increasing the mutual velocity of the condensates results in sharpening the resonance
condition and substantial depopulation of the modes with $p \not \approx Q$.

Let us now consider the situation of two counter-propagating plane matter waves within our approximation in more detail. 
Substituting (\ref{3}) and (\ref{r:3}) into Hamiltonian
(\ref{ham}) we obtain:
\begin{eqnarray}
H &=& \sum_{\qq} \left(  \frac{\hbar^2 q^2}{2m} a^{\dagger}_{\qq}(t) a_{\qq}(t) + \right. \nonumber \\
  &+& \left. \frac{g \sqrt{n_1 n_2}}{2} \ee^{-\ii \hbar \frac{Q^2}{m} t}
a_\qq^\dagger(t) a_{-\qq}^\dagger(t)
+ h.c \right)
\end{eqnarray}
Clearly, the Hamiltonian of the system conserves momentum: two atoms from colliding
plane waves of momenta $\hbar \QQ$ and $-\hbar \QQ$ scatter into atoms of momenta $\hbar \qq$ and $- \hbar
\qq$, the total momentum remains zero.

The Heisenberg equations of motion for annihilation/creation operators can be solved analytically. The
solution takes the following form: \beq a_\pp(t) = \ee^{-\ii \hbar \frac{Q^2}{2m} t} \left( F_1(t,p)
a_{\pp} (0) - F_2(t,p) a^\dagger_{-\pp}(0) \right) \label{r:sol} \eeq where:
\begin{eqnarray}
   F_1(t,p) &=&\cosh (\sqrt{\Delta(p)}\: t) - \II \hbar \frac{p^2 - Q^2}{2m}
   \frac{\sinh (\sqrt{\Delta(p)}\: t)}{\sqrt{\Delta(p)}} \nonumber \\
   F_2(t,p) &=&\II \frac{g \sqrt{n_1 n_2}}{\hbar} \frac{\sinh\left({\sqrt{\Delta(p)}\:t}\right)}{\sqrt{\Delta(p)}}
\end{eqnarray}
and $\Delta(p) = \left( \frac{g \sqrt{n_1 n_2}}{\hbar} \right)^2 - \left( \hbar \frac{p^2-Q^2}{2m}
\right)^2$. Modes can be divided into two classes according to their evolution. The ones which satisfy the
modified energy conservation law, i.e.\ the kinetic energy difference between the incident and scattered atoms is smaller
then the mean field energy: \beq \left(\hbar^2 \frac{p^2 - Q^2}{2 m}\right)^{\!2} < \left(g \sqrt{n_1
n_2}\right)^2 \label{rezon} \eeq have their occupation growing exponentially with time, 
which is a Bose enhanced process.
This relaxation of the energy conservation has an analog in quantum optics, known as power
broadening \cite{11}. The modes that do not satisfy the resonance condition (\ref{rezon}) oscillate with the population of
order unity.

It follows from (\ref{r:sol}) that during the dynamical evolution the state of the system remains a tensor product of
states from 2-mode subspaces. More specifically, if at $t=0$ only modes $\QQ$ and $-\QQ$ were populated, each by
$N/2$ atoms, then at subsequent times the state of the system is of the form:
\beq
 \left|N/2, N/2 \rangle_{\QQ} \right. \otimes
\left( \bigotimes_{\qq\neq\QQ} |\psi(t)\rangle_{\qq} \right) \eeq where \beq |\psi(t)\rangle_{\qq} =
\frac{1}{F_1(t,q)} \sum_{k=0}^{\infty} (-1)^k \left( \frac{F_2(t,q)}{F^*_1(t,q)} \right)^{\!k} |k,k\rangle_\qq
\eeq and $|k,k\rangle_\qq$ denotes a state with $k$ atoms of momentum $\hbar \qq$ and $k$ atoms of momentum
$-\hbar \qq$. Hence, though the state of the system remains pure during dynamics, the statistics of the scattered
atoms is geometric (the same as in a thermal state, which is mixed). Of course our solution is just an approximation to the
particle conserving one, in which the scattered atoms would be entangled with the highly occupied modes with
suitable number of particles missing.

To gain deeper insight into statistical properties of scattered atoms we calculate the
$g^{(2)}(\pp,\qq)$ correlation function: \beq g^{(2)}(\pp,\qq) = \frac{\langle a^\dagger_\pp(t)
a^\dagger_\qq(t) a_\pp(t) a_\qq(t) \rangle}{\langle a^\dagger_\pp(t) a_\pp(t) \rangle \; \langle
a^\dagger_\qq(t) a_\qq(t) \rangle} \eeq Again under the assumption that at $t=0$ all
modes except the $\pm \QQ$ are empty the correlation function becomes: \beq g^{(2)}(\pp,\qq) =  1+
\frac{|F_1(t,p)|^2}{|F_{2}(t,p)|^2} \delta_{\pp,-\qq} + \delta_{\pp,\qq} \eeq We have atom bunching
($g^{(2)}>1$) for two cases: $\pp=-\qq$, which is a result of the momentum conservation law and $\pp=\qq$, which
reflects the Bose enhancement.
Thus, it is expected that
during the collision,
scattered atoms will come out
in the form of pairs of
spikes, randomly placed at the
$p =Q$ circle. Only after the
averaging over many shots the
isotropy will be recovered.

To get closer to reality, let us analyze the situation of two atomic wavepackets colliding. At $t=0$ the
wavepackets envelopes are $f_{\pm \QQ}(\rr)$ and move with momenta $\pm \hbar \QQ$.
Then, at time $t$ the wavefunction of each of the packets 
takes the form: \beq \psi_{\pm \QQ}(\rr,t) = L^{-3} \sum_{\kk}
\tilde{f}_{\pm \QQ}(\kk \pm \QQ) \ee^{-\ii \kk \rr} \ee^{-\ii \hbar \frac{k^2}{2m} t}  \eeq
where $\tilde{f}_{\pm \QQ}(\kk)$ 
denotes the Fourier transform of $f_{\pm \QQ}(\rr)$.
We are using wavefunctions rather than a density matrix since we are dealing with macroscipically populated matter waves.
And even further simplification is made by replacing GPE solution by a free evolution of both colliding wavepackets.
We have therefore entirely neglected the mean field energy in comparison with the kinetic energy of the center of 
mass motion.

Hence, the hamiltionian (\ref{ham}) can be written as:
\begin{eqnarray}
H&=& \sum_{\qq} \left(  \frac{\hbar^2 q^2}{2m} a^{\dagger}_{\qq}(t) a_{\qq}(t) +
       \frac{g}{2 L^6} \sum_{\kk_1, \kk_2} \ee^{-\ii \hbar \frac{k_1^2+k_2^2}{2m} t} \right. \nonumber \\
    & &  \tilde{f}_{\QQ}(\kk_1-\QQ) \tilde{f}_{-\QQ}(\kk_2+\QQ)
      \left.
      a^\dagger_{\qq} a^\dagger_{\kk_1+\kk_2-\qq}
          \right)
\end{eqnarray}
In general the resulting Heisenberg equations for annihilation/creation operators cannot be solved analytically,
but can be treated perturbatively. The first order in the coupling constant $g$ gives the following result:
\begin{eqnarray}
& &  a_\pp(t) = \ee^{-\ii \hbar \frac{p^2}{2m} t} \left[
a_\pp(0) - L^{-6} \frac{\II g}{\hbar} \int_0^t d\tau \; \ee^{\ii\hbar \frac{p^2-Q^2}{m}\tau}
\right. \nonumber \\
& & \left.
\sum_{\uu,\vv} \tilde{f}_{\QQ}(\uu) \tilde{f}_{-\QQ} (\vv) \ee^{-\ii\hbar\frac{\pp(\uu+\vv)+\QQ(\uu-\vv) - \uu\vv}{m}\tau}
     a_{\uu+\vv-\pp}^\dagger(0) \right]
     \label{heisen}
\end{eqnarray}
To evaluate total spontaneous emission losses, one has to consider the quantity: 
\beq 
{\cal{S}}(t) = \sum_{\pp} \langle a_\pp^\dagger(t) a_\pp(t) \rangle \label{r:s} 
\eeq
where the average is taken over the vacuum.

To calculate (\ref{r:s}) we will make further
specifications. We will consider Bragg splitting of a condensate (sometimes called half-collision), which is easier
to realize in experiment than a full collision between two condensates \cite{5}. 
We assume that the initial wavepackets are the Gaussians: \beq
f_{\QQ}(\rr) = f_{-\QQ}(\rr) = \kappa \exp\left[-\frac{1}{2} \left( \frac{x}{\sigma_x} + \frac{y}{\sigma_y}
	  + \frac{z}{\sigma_z} \right)\right] \eeq 
where $\kappa$ fulfills the normalization condition (the collision is along the $x$-axis). 
We will also assume non-dispersive evolution of
wavepackets by neglecting the term $\exp(\II \hbar \uu \vv \tau/ m)$ in (\ref{heisen}), which is justified
by the fact that the scattering takes place while the wavepackets overlap, and for this short time
scale the dispersion does not play a substantial role. In other words,
characteristic time scale of the scattering process is assumed to be very short in comparison with both nonlinear
and dispersive time scales.

Within these approximations the total spontaneous emission losses (\ref{r:s}) can be evaluated analytically:
\begin{eqnarray}
& & {\cal{S}}(t) = \sqrt{\frac{2}{\pi^5}} \; \frac{\hbar^2 a^2 N_1 N_2}{m^2 \sigma_x \sigma_y \sigma_z } \int
d^3 p
\int_0^t d \tau \int_0^t d\tau' \nonumber \\
& & \exp \left( 
	  - \frac{\hbar^2 Q^2}{m^2 \sigma_x^2} \left(\tau^2 + \tau'^2\right)
	  - \frac{\hbar^2}{2 m^2} \left( \tau-\tau'\right)^2 \!\!\!\! \sum_{i=\{x,y,z\}} \frac{p_i^2}{\sigma_i^2}  \right) 
\nonumber \\
& & \exp \left(
	  \II \hbar \frac{p^2-Q^2}{m}(\tau-\tau') \right) 
\label{total}
\end{eqnarray}
In fact, one more integral can be performed analytically, but the resulting formula is even more complex. It
is also worth noting that the order of integrations in (\ref{total}) must be kept -- otherwise one can
easily arrive at a divergent expression. 

\begin{figure}
\begin{center}
\epsfig{file=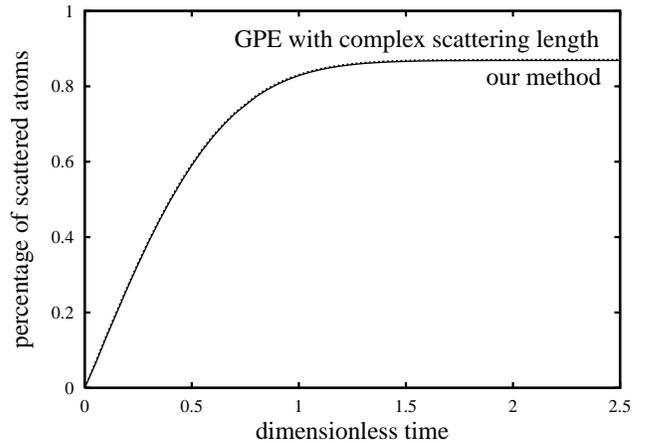,width=8.6cm}
\end{center}
\caption{The growth of spontaneous emission losses relative to the total number of atoms versus time
   in units $m\sigma/(\hbar Q)$.}
\label{rys1}
\end{figure}

In the case of spherical symmetry of the colliding wavepackets ($\sigma_{x,y,z}\!=:\!\sigma$)
the angular distribution of scattered atoms is isotropic and the integral with respect to the
angles of $\pp$ can be performed. 
We have noticed that effectively the total losses do not depend on $|\QQ|$, 
provided that the momentum is sufficiently large to produce
interference fringes in the splitting wavepackets: $Q \sigma \gg 1$. We have checked that in this range total
losses are proportional to: $N_1 N_2 (a/\sigma)^2$. For numerical calculations we assumed
$N_1=N_2=5 \cdot 10^5$ sodium atoms in each wavepacket of width $\sigma = 30 \mu$m and the result is plotted
in Figure \ref{rys1}. For these parameters the value of total spontaneous emission losses is below 1\% of
the total number of particles. The curve is universal if plotted against dimensionless time  $\tilde{t} = t \;
\hbar Q / (m \sigma)$. In this graph we also show for comparison elastic scattering loss from the condensate
calculated from GPE with complex scattering length \cite{9}, where we neglected dispersion and
depletion of the colliding wavepackets. The latter may be described by an analytic formula: ${\cal{S}}(t) = 4 N_1 N_2
(a/\sigma)^2 {\mbox{Erf}}(\tilde{t}\sqrt{2})$ (shown as a dashed line in Figure \ref{rys1}, but the curves are
											   practically indistinguishable).
Both methods are first order in scattering length and agree extremely well.

The influence of the geometry of the colliding wavepackets on the total losses has been investigated for
cylindrically symmetric gaussians. 
For a wide range of the aspect ratio 
the total losses do not depend on the geometry
(provided that the condition $Q \sigma_x \gg 1$ is fulfilled). Similar effect
is predicted by the complex scattering length method.

The method presented here shows extraordinary good agreement with the predictions of the complex scattering 
length technique within the approximations discussed earlier. 
It can be shown (using dynamic equations for non commutative cumulants)
that within first order approximation both methods are essentailly
equivalent \cite{Burnett}.
It would be interesting to go beyond these 
approximations to check the predictions of \cite{9}, we are currently studying the Bose enhancement
for realistic condensate droplets. 
It is worth stressing however that our method not only measures total spontaneous emission
losses during the collison, but also gives insight into
the statistical properties of the scattered atoms.

The recent experiment by MIT group \cite{Ketterle}, measuring the losses in the splitted condensate, was 
performed in different regime than discussed in this article. The measured fraction of collided atoms 
was up to 20\%, which is clearly beyond the first order perturbation theory described above. 
Authors of \cite{Ketterle} argue that 
the increase of the fraction of collided atoms with the number of outcoupled impurities (in our language that is
the total loss versus the number of atoms in one of the packets, keeping the total number of atoms in both 
condensates fixed) is an evidence of collective self-amplification and is due to bosonic enhancement. 
It is striking however that the shape of this dependence (Fig.3 of \cite{Ketterle}) 
resembles the parabola obtained from the formula (\ref{total}).

To summarize: We have proposed a method that allows to calculate the spontaneous emission losses of colliding condensates.
Contrary to earlier attempts \cite{9}, our approach uses wave functions rather than particle densities. The coherence of the
colliding droplets is therefore taken into account. The method is applicable to sufficiently quickly 
moving condensates, i.e.
when the spread of momenta in each condensate is much smaller than their relative momentum 
   and in this regime is essentially equivalent to Bogoliubov approach. For the simplest case of colliding plane waves
we pointed out the thermal-like multiplicity distribution of scattered atoms in each mode and the bunching reflecting the
momentum conservation on one hand and the bosonic enhancement on the other. Similar effects must be also present for the
colliding wavepackets.

R.B. and K.Rz. acknowledge the support of the subsidy from Foundation for Polish Science, while M.T.
acknowledges support from KBN nr 2P03B7819.

\end{document}